\documentclass[runningheads,fleqn]{cl2emult}
\usepackage{makeidx}  % allows index generation
\usepackage{graphicx} % standard LaTeX graphics tool
\usepackage{subeqnar} % subnumbers individual equations
\usepackage{multicol} % used for the two-column index
\usepackage{cropmark} % cropmarks for pages without
\usepackage{phys}     % flushleft layout of math and captions
\makeindex            % used for the subject index
\newcommand{\CD}{\ensuremath{{\cal D}}}

\newcommand{\boldtau}{\mathbf{\tau}}
\newcommand{\boldsigma}{\mathbf{\sigma}}

\begin{document}
%
%AUTHOR_STYLES_AND_DEFINITIONS%%%%%%%%%%%%%%%%%%%%%%%%%%%%%%%
%
%Please reduce your own definitions and macros to an absolute
%minimum since otherwise the editor will find it rather
%strenuous to compile all individual contributions to a
%single book file

%%%%%%%%%%%%%%MMmacros
% bold italic/greek fonts
\font\tenmib=cmmib10
\font\sevenmib=cmmib7
\font\fivemib=cmmib5

\newfam\mibfam \def\mib{\fam\mibfam\tenmib}
	\textfont\mibfam=\tenmib \scriptfont\mibfam=\sevenmib
	\scriptscriptfont\mibfam=\fivemib

% bold math symbol fonts
\font\tenbsy=cmbsy10
\font\sevenbsy=cmbsy7
\font\fivebsy=cmbsy5

\newfam\bsyfam \def\bsy{\fam\bsyfam\tenbsy}
\textfont\bsyfam=\tenbsy \scriptfont\bsyfam=\sevenbsy
	\scriptscriptfont\bsyfam=\fivebsy

%\mathchardef\gamma="010D
%\mathchardef\boldgamma="0\mibfam0D

% \mathchardef\boldalpha="0\mibfam0B
\def\boldalpha{{\mib\mathchar"700B}}
\def\boldbeta{{\mib\mathchar"700C}}
\def\boldgamma{{\mib\mathchar"700D}}
\def\bolddelta{{\mib\mathchar"700E}}
\def\boldepsilon{{\mib\mathchar"700F}}
\def\boldzeta{{\mib\mathchar"7010}}
\def\boldeta{{\mib\mathchar"7011}}
\def\boldtheta{{\mib\mathchar"7012}}
\def\boldiota{{\mib\mathchar"7013}}
\def\boldkappa{{\mib\mathchar"7014}}
\def\boldlambda{{\mib\mathchar"7015}}
\def\boldmu{{\mib\mathchar"7016}}
\def\boldnu{{\mib\mathchar"7017}}
\def\boldxi{{\mib\mathchar"7018}}
\def\boldpi{{\mib\mathchar"7019}}
\def\boldrho{{\mib\mathchar"701A}}
\def\boldsigma{{\bf \sigma}}
\def\boldtau{{\tau}}
\def\boldupsilon{{\mib\mathchar"701D}}
\def\boldphi{{\mib\mathchar"701E}}
\def\boldchi{{\mib\mathchar"701F}}
\def\boldpsi{{\mib\mathchar"7020}}
\def\boldomega{{\mib\mathchar"7021}}
\def\boldvarepsilon{{\mib\mathchar"7022}}
\def\boldvartheta{{\mib\mathchar"7023}}
\def\boldvarpi{{\mib\mathchar"7024}}
\def\boldvarrho{{\mib\mathchar"7025}}
\def\boldvarsigma{{\mib\mathchar"7026}}
\def\boldvarphi{{\mib\mathchar"7027}}

% capital bold greek letters also available as {\mib\Gamma}, etc.
\def\boldGamma{{\mib\mathchar"7000}}
\def\boldDelta{{\mib\mathchar"7001}}
\def\boldTheta{{\mib\mathchar"7002}}
\def\boldLambda{{\mib\mathchar"7003}}
\def\boldXi{{\mib\mathchar"7004}}
\def\boldPi{{\mib\mathchar"7005}}
\def\boldSigma{{\mib\mathchar"7006}}
\def\boldUpsilon{{\mib\mathchar"7007}}
\def\boldPhi{{\mib\mathchar"7008}}
\def\boldPsi{{\mib\mathchar"7009}}
\def\boldOmega{{\mib\mathchar"700A}}

\def\boldell{{\mib\mathchar"7060}}

\def\boldaleph{{\bsy\mathchar"7040}}
\def\boldnabla{{\bsy\mathchar"7072}}
\def\boldcdot{{\bsy\mathchar"7001}}
\def\boldtimes{{\bsy\mathchar"7002}}
\def\boldast{{\bsy\mathchar"7003}}

%% boldface "cal"

\def\AB{{\bsy\mathchar"7041}}
\def\BB{{\bsy\mathchar"7042}}
\def\CB{{\bsy\mathchar"7043}}
\def\DB{{\bsy\mathchar"7044}}
\def\EB{{\bsy\mathchar"7045}}
\def\FB{{\bsy\mathchar"7046}}
\def\GB{{\bsy\mathchar"7047}}
\def\HB{{\bsy\mathchar"7048}}
\def\IB{{\bsy\mathchar"7049}}
\def\JB{{\bsy\mathchar"704A}}
\def\KB{{\bsy\mathchar"704B}}
\def\LB{{\bsy\mathchar"704C}}
\def\MB{{\bsy\mathchar"704D}}
\def\NB{{\bsy\mathchar"704E}}
\def\OB{{\bsy\mathchar"704F}}
\def\PB{{\bsy\mathchar"7050}}
\def\QB{{\bsy\mathchar"7051}}
\def\RB{{\bsy\mathchar"7052}}
\def\SB{{\bsy\mathchar"7053}}
\def\TB{{\bsy\mathchar"7054}}
\def\UB{{\bsy\mathchar"7055}}
\def\VB{{\bsy\mathchar"7056}}
\def\WB{{\bsy\mathchar"7057}}
\def\XB{{\bsy\mathchar"7058}}
\def\YB{{\bsy\mathchar"7059}}
\def\ZB{{\bsy\mathchar"705A}}

%%
%%  miscellaneous Math Macros
%%
%%
\def\Bp#1{\bigl(#1\bigr)}
\def\BP#1{\Bigl(#1\Bigr)}
\def\Bpp#1{\biggl(#1\biggr)}
\def\BpP#1{\Biggl(#1\Biggr)}
\def\Bb#1{\bigl[#1\bigr]}
\def\BB#1{\Bigl[#1\Bigr]}
\def\Bbb#1{\biggl[#1\biggr]}
\def\BbB#1{\Biggl[#1\Biggr]}
\def\Bbr#1{\bigl\{#1\bigr\}}
\def\BBr#1{\Bigl\{#1\Bigr\}}
\def\Bbbr#1{\biggl\{#1\biggr\}}
\def\BbBr#1{\Biggl\{#1\Biggr\}}
\def\Er#1{e^{#1}}
\def\Erp#1#2{e^{#1(#2)}}
\def\ErBp#1#2{e^{#1\Bp{#2}}}
\def\ErBP#1#2{e^{#1\BP{#2}}}
\def\ErBpp#1#2{e^{#1\Bpp{#2}}}
\def\ErBpP#1#2{e^{#1\BpP{#2}}}
\def\Erb#1#2{e^{#1[#2]}}
\def\ErBb#1#2{e^{#1\Bb{#2}}}
\def\ErBB#1#2{e^{#1\BB{#2}}}
\def\ErBbb#1#2{e^{#1\Bbb{#2}}}
\def\ErBbB#1#2{e^{#1\BbB{#2}}}
\def\Erbr#1#2{e^{#1\{#2\}}}
\def\ErBbr#1#2{e^{#1\Bbr{#2}}}
\def\ErBBr#1#2{e^{#1\BBr{#2}}}
\def\ErBbbr#1#2{e^{#1\Bbbr{#2}}}
\def\ErBbBr#1#2{e^{#1\BbBr{#2}}}
\def\Exp#1{\exp(#1)}
\def\ExBp#1{\exp\Bp{#1}}
\def\ExBP#1{\exp\BP{#1}}
\def\ExBpp#1{\exp\Bpp{#1}}
\def\ExBpP#1{\exp\BpP{#1}}
\def\Exb#1{\exp[#1]}
\def\ExBb#1{\exp\Bb{#1}}
\def\ExBB#1{\exp\BB{#1}}
\def\ExBbb#1{\exp\Bbb{#1}}
\def\ExBbB#1{\exp\BbB{#1}}
\def\Exbr#1{\exp\{#1\}}
\def\ExBbr#1{\exp\Bbr{#1}}
\def\ExBBr#1{\exp\BBr{#1}}
\def\ExBbbr#1{\exp\Bbbr{#1}}
\def\ExBbBr#1{\exp\BbBr{#1}}
\def\Pr{^{\prime\hskip-0.07em\relax}}
\def\Spr{^{\,\prime\hskip-0.07em\relax}}
\def\Dpr{^{\prime\hskip-0.09em\relax\prime\hskip-0.07em\relax}}
\def\Tpr{^{\prime\hskip-0.09em\relax\prime\hskip-0.07em\relax
\prime\hskip-0.07em\relax}}
\def\Ylm#1,#2,#3{{\rm Y}_{#1#2}(#3)}
\def\Clm#1,#2,#3{{\rm C}_{#1#2}(#3)}
\def\Iylm#1,#2,#3{{\cal Y}_{#1#2}(#3)}
\def\IylmC#1,#2,#3{{\cal Y}_{#1#2}^\ast(#3)}
\def\Iclm#1,#2,#3{{\cal C}_{#1#2}(#3)}
\def\IclmC#1,#2,#3{{\cal C}_{#1#2}^\ast(#3)}
\def\Jm#1,#2{{\rm J}_{#1}(#2)}
\def\jl#1,#2{j_{#1}(#2)}
\def\Jl#1,#2,#3{{\cal J}_{#1}(#2,#3)}
\def\Nm#1,#2{{\rm N}_{#1}(#2)}
\def\nl#1,#2{n_{#1}(#2)}
\def\Nl#1,#2,#3{{\cal N}_{#1}(#2,#3)}
\def\Hm#1,#2{{\rm H}_{#1}(#2)}
\def\hl#1,#2{h_{#1}(#2)}
\def\Hl#1,#2,#3{{\cal H}_{#1}(#2,#3)}
\def\Kl#1,#2,#3{{\cal K}_{#1}(#2,#3)}
\def\Subpa{_{\parallel}}
\def\Subpe{_{\perp}}
\def\Xvpa{\vec x\Subpa}
\def\Kg{\vec k+\vec G}
\def\Ekgx{e^{i(\Kg)\cdot\Xvpa}}
\def\Tpt{\vec\tau-\vec\tau\Spr}
\def\Tptr{\Tpt+\vec R}
\def\Xtptr{\vec x-(\Tptr)}
\def\Ekgtc{e^{-i(\Kg)\cdot(\Tpt)\Subpa}}
\def\Ekgxc{e^{-i(\Kg)\cdot\vec x\Subpa}}
\def\Fpjl{{1\over4\pi\Jl\ell,\kappa,x}}
\def\Incc{\int\hskip-0.3em d{\hat x}\ \IclmC\ell,m,{\hat x}}
\def\Icdj{\Fpjl\Incc}
\def\Blsu#1,#2{\sum_{#1}e^{i\vec{#2}\cdot\vec{#1}}}
\def\Excl{\bigl(1-\delta(R,0)\delta(\tau\Pr\tau)\bigr)}
\def\Inxt{\int_{0(C)}^{\infty}\hskip-0.3em dt\ }
\def\Inxa{\int_{0(C)}^{\xi/2}\hskip-0.3em dt\ }
\def\Inxb{\int_{\xi/2}^{\infty}\hskip-0.3em dt\ }
\def\Torp{{2\over\sqrt{\pi}}}
\def\Qquad{\qquad\qquad}
\def\QQuad{\Qquad\Qquad}
\def\QQUad{\QQuad\QQuad}
\def\Arg{{\rm arg}}
\def\Conjg#1{{#1}^{\boldast}}

% for left justified displayed equations
\def\leftdisplay#1$${\leftline{\indent$\displaystyle{#1}$}$$}

% Norm
\def\Norm#1{|{#1}|}

% Rnorm
\def\Rnorm#1{|\hskip-0.14em|{#1}|\hskip-0.14em|}

% Grad
\def\Grad{\vec\nabla}

% Fac
\def\Fac{\kern-.2pt !}
% Dblfac
\def\Dblfac{!\hskip-0.10em!}
% doublebar
\def\doublebar{\big|\kern-.5pt\big|}
\def\Doublebar{\Big|\kern-.5pt\Big|}

% \Half
\def\Half{\hskip0.05em^1\hskip-0.14em/\hskip-0.10em_2}
\def\Frac#1#2{\hskip0.05em^{#1}\hskip-0.14em/\hskip-0.10em_{#2}}
\def\Sfrac#1#2{{\kern-.15em {#1 \over #2}}}

% \erf, \erfc
\def\erf{{\rm erf}}
\def\erfc{{\rm erfc}}

% \sgn
\def\sgn{{\rm sgn}}

% small minus and plus
\def\sminus{{\cal\mathchar"7000}}
\def\splus{{\rm\mathchar"702B}}

% vector coupling coefficients
\def\vccof#1#2#3#4#5#6{
	\left( #1~#2~#3~#4 \big| #1~#3~#5~#6 \right)
}
%
% Wigner coefficients.
%
\def\wignercof#1#2#3#4#5#6{
	\left( \matrix{ #1 & #2 & #3 \cr #4 & #5 & #6 \cr } \right)}
\def\wccof#1,#2,#3,#4,#5,#6;{
	(-)^{#3\sminus #1\sminus #6}
	\bigl( 2#5\splus 1 \bigr)^{\Sfrac12}
	\wignercof{#1}{#3}{#5}{#2}{#4}{\sminus #6}
}
\def\LM#1#2{#1\thinspace #2}
%
% reduced matrix elements
%
\def\reducedme#1#2#3{\bigl( #1 \doublebar #2 \doublebar #3 \bigr)}
\def\Reducedme#1#2#3{\Bigl( #1 \Doublebar #2 \Doublebar #3 \Bigr)}
%
% Gradient coefficients.
\def\gradcof#1#2#3#4{ \bigl( #1\big|#2~#3~#4 \bigr)}
% six j symbol
\def\sixj#1#2#3#4#5#6{
	\left\lbrace \matrix{ #1 & #2 & #3 \cr #4 & #5 & #6 \cr } \right\rbrace}

% (my) gaunt coefficient
\def\gauntcof#1#2#3#4#5#6{
	\G\bigl(\thinspace
		#1,\thinspace
		#2;\thinspace
		#3,\thinspace
		#4;\thinspace
		#5,\thinspace
		#6\thinspace
	\bigr)
}
% (plain) gaunt coefficient
\def\pgauntcof#1#2#3{\RG\bigl(#1\thinspace;\thinspace #2\thinspace;\thinspace #3\thinspace\bigr)}

% vector spherical harmonic
\def\vshy#1,#2,#3;#4;{{\mib Y}_{#1\thinspace #2\thinspace #3\thinspace}(#4)}
\def\vshiy#1,#2,#3;#4;{{\YB}_{#1\thinspace #2\thinspace #3\thinspace}(#4)}
\def\vshc#1,#2,#3;#4;{{\mib C}_{#1\thinspace #2\thinspace #3\thinspace}(#4)}
\def\vshic#1,#2,#3;#4;{{\CB}_{#1\thinspace #2\thinspace #3\thinspace}(#4)}

% cross produce \Cross
\def\Cross{\land}

% Vector symbols
%
% bold face roman letters
\def\bfrom#1{{\bf #1}}
\def\bvc#1{{\bf #1}}
\def\uvc#1{\hat\bvc #1}

% cal math letters

\def\A{{\cal A}}
\def\B{{\cal B}}
\def\C{{\cal C}}
\def\D{{\cal D}}
\def\E{{\cal E}}
\def\F{{\cal F}}
\def\G{{\cal G}}
\def\H{{\cal H}}
\def\I{{\cal I}}
\def\J{{\cal J}}
\def\K{{\cal K}}
\def\L{{\cal L}}
\def\M{{\cal M}}
\def\N{{\cal N}}
\def\O{{\cal O}}
\def\P{{\cal P}}
\def\Q{{\cal Q}}
\def\R{{\cal R}}
\def\S{{\cal S}}
\def\T{{\cal T}}
\def\U{{\cal U}}
\def\V{{\cal V}}
\def\W{{\cal W}}
\def\X{{\cal X}}
\def\Y{{\cal Y}}
\def\Z{{\cal Z}}

% Roman letters

\def\RA{{\rm A}}
\def\RB{{\rm B}}
\def\RC{{\rm C}}
\def\RD{{\rm D}}
\def\RE{{\rm E}}
\def\RF{{\rm F}}
\def\RG{{\rm G}}
\def\RH{{\rm H}}
\def\RI{{\rm I}}
\def\RJ{{\rm J}}
\def\RK{{\rm K}}
\def\RL{{\rm L}}
\def\RM{{\rm M}}
\def\RN{{\rm N}}
\def\RO{{\rm O}}
\def\RP{{\rm P}}
\def\RQ{{\rm Q}}
\def\RR{{\rm R}}
\def\RS{{\rm S}}
\def\RT{{\rm T}}
\def\RU{{\rm U}}
\def\RV{{\rm V}}
\def\RW{{\rm W}}
\def\RX{{\rm X}}
\def\RY{{\rm Y}}
\def\RZ{{\rm Z}}

%
% a box
%
\def\littlebox{%
      \hbox{%
      \vrule\vbox{\hrule width .7em\kern1.5ex%
      \hrule width .7em}\vrule}%
}

\def\Dnx#1#2{d^{#1}\kern-1.2pt#2\ }

\def\bra#1{\bigl\langle #1 \big|}
\def\Bra#1{\Bigl\langle #1 \Big|}
\def\braa#1{\biggl\langle #1 \bigg|}
\def\Braa#1{\Biggl\langle #1 \Bigg|}

\def\ket#1{\big| #1 \bigr\rangle}
\def\Ket#1{\Big| #1 \Bigr\rangle}
\def\kett#1{\bigg| #1 \biggr\rangle}
\def\Kett#1{\Bigg| #1 \Biggr\rangle}

\def\braket#1#2{\bigl\langle #1 \big| #2 \bigr\rangle}
\def\Braket#1#2{\Bigl\langle #1 \Big| #2 \Bigr\rangle}
\def\brakett#1#2{\biggl\langle #1 \bigg| #2 \biggr\rangle}
\def\Brakett#1#2{\Biggl\langle #1 \Bigg| #2 \Biggr\rangle}

\def\avg#1{\big\langle #1 \big\rangle}
\def\Avg#1{\Big\langle #1 \Big\rangle}
\def\avgg#1{\bigg\langle #1 \bigg\rangle}
\def\Avgg#1{\Bigg\langle #1 \Bigg\rangle}

\def\Tr{{\rm T\kern-1.5pt r\kern 1.5pt}}
\def\grad{\boldnabla}
\def\laplacian{\nabla^2}

%
% Hankel error function
%
\def\hrfc{{\rm hrfc}}
%
% left aligned, double tabbed, numbered equations (maybe)
%
%\catcode`\@=11
%\def\eqldtno#1{\displ@y \tabskip=0pt
%	\halign to\displaywidth{$\@lign\displaystyle{##}$\tabskip=0pt
%	&$\@lign\displaystyle{{}##}$\hfil\tabskip=0pt
%	&$\@lign\displaystyle{{}##}$\hfil\tabskip=0pt
%	&\llap{$\@lign##$}\tabskip=0pt\crcr
%	#1\crcr}}
%\catcode`\@=12
\catcode`\@=11
\def\eqlstno#1{\displ@y \tabskip=0pt
	\halign to\displaywidth{$\@lign\displaystyle{##}$\tabskip=0pt
	&$\@lign\displaystyle{{}##}$\hfil\tabskip=\centering
	&\llap{$\@lign##$}\tabskip=0pt\crcr
	#1\crcr}}
\catcode`\@=12

\def\pFq#1,#2;#3,#4;#5;{\, _{#1}F_{#2}\Bigl(
	{\textstyle {#3 \atop #4} } \Big| #5 \Bigr)}

\def\pGq#1,#2;#3,#4;#5,#6;#7;{\,G^{{#1},{#2}}_{{#3},{#4}}\Bigl(
	{#7} \Big| {\textstyle {{#5} \atop {#6}}} \Bigr)}

% Derivatives
\def\PDir#1{{\partial\over\partial #1}}

% Logic
\def\suchthat{s\kern-1.5pt t\kern1.5pt}

%% Group symmetry
\def\sgelem#1#2{\bigl( #1 \big| #2 \bigr)}
\def\Sgelem#1#2{\Bigl( #1 \Big| #2 \Bigr)}
\def\sggelem#1#2{\biggl( #1 \bigg| #2 \biggr)}
\def\Sggelem#1#2{\Biggl( #1 \Bigg| #2 \Biggr)}

%% convolution
\def\Conv{{\rm\textstyle con}}
%% Solid harmonics
\def\solY{X} \def\isolY{{\cal X}}
%% mod
\def\Mod#1#2{#1{\rm mod}_{#2}}
%% bold letters
\def\r{{\bf r}}

\title*{ Excited States Calculated by Means of the Linear Muffin-Tin Orbital
Method  }
%\protect\newline 
%
%
\toctitle{Excited States 
%by means of the Linear muffin-tin orbital method 
}
%\protect\newline in the Particle Deflection Plane}
% allows explicit linebreak for the table of content
%
%
\titlerunning{Excited states}
% allows abbreviation of title, if the full title is too long
% to fit in the running head
%
\author{M. Alouani\inst{1}
\and J. M. Wills\inst{2} }
\authorrunning{M. Alouani and J. M. Wills}
% if there are more than two authors,
% please abbreviate author list for running head
%
%
\institute{IPCMS, Universit\'e Louis Pasteur, 23 Rue du Loess, 67037 
Strasbourg, France
\and Los Alamos National Laboratory, Los Alamos, NM 87545, USA}

\maketitle              % typesets the title of the contribution

\begin{abstract}
The most popular electronic structure method, the linear muffin-tin
orbital method (LMTO), in its full-potential (FP) and relativistic 
forms has been extended to calculate the spectroscopic properties of 
materials form first principles, i.e, optical spectra, x-ray magnetic
circular dichroism (XMCD) and magneto-optical kerr effect (MOKE).  
The paper describes an  overview of the FP-LMTO  basis set 
and the calculation of the momentum matrix elements. 
Some applications concerning  the computation  of  optical properties of 
semiconductors and  XMCD spectra of transition metal alloys are
reviewed. 
\end{abstract}

\section{Introduction}
%input introduction.tex
The density functional theory (DFT) of Hohenberg, Kohn, and Sham is the method
of choice for describing the ground-state properties of materials \cite{KS}. 
However, in the initial derivation  of the DFT, the eigenvalues are 
Lagrange multipliers introduced  to orthogonalize
the eigenvectors, which in their turn are used to compute the total 
energy and the charge density. In this formulation the eigenvalues have
therefore no physical meaning and should not be considered as 
excited states. Nevertheless, the DFT in the local density approximation (LDA) 
or in its spin resolved local density formulation (LSDA), 
has been used successfully to compute the excited states,
namely, optical and magneto-optical properties, x-ray absorption 
and magnetic dichroism spectra.

The LDA or LSDA were indeed intended to compute  the ground-state properties
of materials, and their  use during the last two decades has produced an 
excellent track record in the computation   of these properties 
for a wide variety of materials, ranging from simple metals to 
complex semiconductor superlattices.
However, it is now believed that the DFT can do more than computing the ground
state properties. This is because  the Kohn--Sham equations could be viewed as deriving 
form a simplified quasi-particle (QP) theory where the self-energy is local and 
time averaged, i.e.,    $\Sigma({\bf r}, {\bf r^\prime}, t) \approx V_{xc} ({\bf r})
\delta ({\bf r} -  {\bf r^\prime} ) \delta (t)$, here $V_{xc} ({\bf r})$
is the local exchange and correlation potential as, for example,
parameterized by Von Barth and
Hedin \cite{vbarth}.  Viewed in this way, the KS  eigenvalues are
then approximate QP energies and could be compared to experimental data.
This argument is supported by quasiparticle calculations 
within  the so called GW approximation of 
Hedin \cite{hedin} showing that the valence QP energies
 of semiconductors are in good agreement with these obtained using LDA, and the 
conduction QP energies differ by approximately a rigid energy 
shift \cite{hl,gss}.  In the literature this shift is often called 
``scissors-operator'' shift \cite{zachary}.

In the last few years spectroscopy is  becoming the standard
tool for measuring excited states of materials.  
Its owes its  impressive advances  to the availability of synchrotron tunable
highly polarized radiation. 
In particular, the measurement of optical, magneto-optical
properties as well as magnetic x-ray dichroism are now becoming routine tasks
for probing the structural and  magnetic properties of materials.  
Considerable attention has been focused on transition-metal surfaces and 
and thin films due to their novel physical properties different from
that of bulk materials and due to potential industrial applications
such as magneto-optical recording, sensors, or  technology based
on giant magneto-resistance. In this respect, theory is falling far behind
experiment and it is becoming hard to give a basic interpretation  of 
experimental data.

This paper, which is far from being  a review paper about calculated
excited states, tries to bridge the gap between experiment and theory
by  
describing  a rather quantitative  method for computing excited states 
of materials. 
This method uses  the local density approximation and the linear muffin-tin
orbital (LMTO) method. In the first part of this paper we introduce the 
density functional theory and the local density approximation and justify 
the use of LDA eigenvalues as approximate excited states
and relate  them to quasiparticle energies. In the second part we 
describe the construction of  the LMTO basis set within an 
all-electron full-potential approach \cite{jwarticle,oka} which will be
used to determine the momentum matrix elements. 
We devote the third part  to the determination of the momentum  matrix 
elements.
In the first part of the application section we 
 present some examples of computation  of semiconductors 
optical spectra \cite{logoetal,aw}, and leave out the
optical properties of metals and magneto-optical properties of materials and 
refer the reader to Ref.
\cite{delin,ahuja,callaway,uspenski,akk,antonov,halilovyu,openeer,wierman}. 
In the second part of the applications we show some examples of   
x-ray magnetic dichroism calculations \cite{aww,grange,iosif}. 

\section{Density Functional Theory}
%input df.tex
The density functional method of Hohenberg and Kohn \cite{KS} 
which states that the
ground state total energy of a system of N interacting electrons in an
external potential $V_{\rm{ext}}$ is a functional of the electron density $\rho (\r
)$ does not provide an analytical form of the functional \cite{KS}.  
This method remains
numerically intractable without the Kohn and Sham introduction of the
so called local density approximation \cite{KS} in which the exchange and
correlation
functional $E_{xc}\{n\}$ appearing in the total energy:

\begin{eqnarray}
\lefteqn{ E\{n\} = T\{n\}
      + {e^2\over2} \int d^3r \int d^3r\Pr
            {n(\bvc r)n(\bvc r\Pr)\over |\bvc r-\bvc r\Pr|} 
      + E_{xc}\{n\}} \nonumber \\ 
      && {} + \int d^3r\ V_{\rm{ext}}(\bvc r) n(\bvc r)
      + E(V_{\rm{ext}})
\end{eqnarray}

is  given by $E_{xc}\{n\} = \int d^3r\ \epsilon_{xc}\bigl(n(\bvc r)\bigr)
n(\bvc r)$ where $\epsilon_{xc}$ is the exchange-correlation energy of a
uniform electron gas of density $n$. Thus, Kohn and Sham constructed a
set of self-consistent single-particle equations:

\begin{equation}
\Bigl(- \nabla^2 + {\delta\over\delta n}\bigl(E - T\bigr) \Bigr) \psi(\bvc r)
      = e_i \psi_i(\bvc r)
\end{equation}

where the density $n(\bvc r)$ is given by:

\begin{equation}
      n(\bvc r)
      = \sum_{i} \theta(e_i<E_F) \psi_i(\bvc r) \psi_i^{\dagger}(\bvc r)
\end{equation}

\noindent
and
\begin{equation}
      V_{\rm{ext}}(\bvc r)
      = - e^2 \sum_{R\tau} {Z_\tau \over |\bvc r-\boldtau-\bvc R|}
\end{equation}

\begin{equation}
      E(V_{\rm{ext}})
      = e^2 \sum_{\tau R}\sum_{\tau\Pr R\Pr}
               \bigl(1 - \delta(R,R\Pr)\delta(\tau,\tau\Pr)\bigr)
               {Z_{\tau} Z_{\tau\Pr} \over |\boldtau+\bvc
R-\boldtau\Pr-\bvc R\Pr|}
\end{equation}

Instead of the true kinetic energy of the electron gas, Kohn and Sham used
the homogeneous electron kinetic energy:

\begin{equation}
      \bar T
      \equiv \sum_{i}\theta(e_i<E_{\rm{F}})\ \int d^3r\ \psi_i^\dagger
(-\nabla^2) \psi_i
\end{equation}

This use of homogeneous-electron kinetic energy in the Kohn--Sham equations
redefined the exchange-correlation function to be:

\begin{equation}
      \bar E_{xc}\{n\} \equiv \bigl(E_{xc}\{n\} + T\{n\} - \bar T \bigr)
      =  \int d^3r\ \epsilon_{xc}\bigl(n(\bvc r)\bigr) n(\bvc r)
\end{equation}

It is then crucial to use a good basis-set for the description of 
the electronic structure  of realistic 
systems. The augmented plane wave \cite{apw} (APW), and the Korringa-Kohn-Rostoker
\cite{kkr} (KKR)
methods can be used, in principle, to solve exactly the Kohn--Sham equations, 
however these methods are numerically involved and their  linearization, 
introduced by Andersen is much preferable. Andersen linearization, has
not only made the techniques for solving the band-structure problem 
transparent by reducing it essentially to  the diagonalization of 
one-electron Hamiltonian, and  cuts the cost of
computation  by at least one order of magnitude. The linearized versions
of these two powerful methods are  the linear
augmented plane wave (LAPW) and linear muffin-tin orbital (LMTO) methods, 
respectively \cite{oka}.

In this paper, we will only use the LMTO method to 
study  excited states of solids. The reason for this choice is 
that the  LMTO method is the mostly used method in computational
electronic structure.  This is due primarily to the use of atomic-sphere 
approximation (ASA) which made the LMTO method run fast even on today's 
cheap personal computers. Due to this reduced computational cost,
the LMTO ASA method became the method of   choice of researchers without 
access to supercomputers.

\section{Quasiparticle Theory and Local-Density Approximation Link}
%input gw.tex
The quasiparticle (QP) electronic structure of an interacting many-body system
is described by the single-particle eigenstates resulting from the
interaction of this single particle with the many-body electron gas of
the system. The single-particle eigenstate energies are the results of 
solving a Schroedinger like equation containing the non-local and
energy-dependent self-energy instead of the exchange-correlation potential
appearing in Kohn--Sham like equations:

\begin{displaymath}
(T + V_H + V_{\rm{ext}})\Psi({\bf r}) + \int d^3r' \Sigma({\bf r},{\bf r'},E)
\Psi({\bf r'}) = E\Psi({\bf r}).
\end{displaymath}

Thus the self-energy $\Sigma$ contains all many-body effects. Almost all
ab-initio QP studies were performed within the so-called $GW$ approximation,
where the self-energy $\Sigma$ is calculated within Hedin's $GW$
approximation. This method consists of approximating the self-energy
as the convolution of the LDA  self-consistent  Green function $G$ and the
screened coulomb interaction $W$ within  the random-phase approximation. 
The QP eigenvalues are often obtained 
 using  first-order perturbation theory 
starting from LDA eigenvalues and eigenvectors \cite{hl,godby}. 
Although there are 
early calculations starting from Hartree-Fock \cite{strinati} or
tight-binding \cite{stern} methods. Nevertheless, the best results are based
on a LDA starting point \cite{hl,godby,germans,arya,willy}.  

Thus the $GW$ predicted optical
excitations energies of semiconductors are within 0.1 eV form the
experimental results and the surprizing fact is that the 
QP wave functions are almost identical to these
produced within the LDA \cite{hl} (the wave function overlap is more
than 99\%).  For a general review of GW calculations see the 
review by  Araysetianwan and Gunnarsson \cite{arya} or by Aulbur, J\"onsson 
and Wilkins \cite{willy}.

It is clear that the quasiparticle Schroedinger equation resembles to 
the Kohn--Sham equation. Both equations describe a fictitious  electron
moving in a effective potential. The difference is that the self-energy is
nonlocal and energy dependent whereas the LDA potential is local and 
averaged over time. This resemblance can be further pushed by noticing
that the DFT can be used to obtain excitation energies. For example, the
ionization energy, $I$,  and the electron affinity, $A$, are difference
between ground state energies:

\begin{displaymath}
I = E (N-1) - E(N), \hskip 1truecm {\rm and} \hskip 1truecm  A = E(N) - E(N+1) 
\end{displaymath} 

Were $N$ is the number of electrons of the system. And since the DFT gives
the correct ground state energies it should produce, in principle, the
correct ionization and electron affinity energies. For metals, the
addition or removal of an electron from the system costs the same energy,
and hence the ionization energy is equal to the electron affinity. For
insulators, the energy gap makes all the difference and hence breaks this
symmetry. Thus the energy band gap is given by:

$E_g = I-A = E(N+1) + E(N-1) - 2 E( N)$

In practice, however, the calculation is often obtained within the LDA and
the energy band gap is calculated as the difference between the lowest 
conduction band and the highest valence band. It was shown by Sham and 
Schl\"uter \cite{shams} and Perdew and Levy \cite{plevy} that the calculated 
energy gap differ from
the true band gap by an amount $\Delta$ even when the DFT is used without
the LDA.   The $\Delta$ value could range from $50\%$ in the case of silicon
to $100\%$ in the case of germanium. For most of the semiconductors,
the  $GW$ calculations show that the LDA eigenvalues differ form the $GW$
quasiparticle energy by a constant $\Delta$ which is almost independent
of the $\bf k$-point. This finding is important and shows that the 
LDA eigenvalues  have some meaning and could be used to calculate 
excited states. So as stated in the introduction,  
the Kohn--Sham equations could be viewed as deriving
form a simplified quasi-particle (QP) theory where the self-energy is made
local and time averaged, i.e.,    
$\Sigma({\bf r}, {\bf r^\prime}, t) \approx V_{xc} ({\bf r})
\delta ({\bf r} -  {\bf r^\prime} ) \delta (t)$. This approximation is
certainly good for  metals where we have a good data base for 
excited state calculated within the 
LDA \cite{delin,ahuja,callaway,uspenski,akk,antonov,halilovyu,openeer,wierman} 
and where the agreement with 
experiment is good. For semiconductors, this approximation is not bad
either, provided we know the value of the discontinuity of the exchange
and correlation. Usually, this value is  provided by $GW$ calculations or by
experiment. 
\section{The Full-Potential LMTO Basis Set}
%input basis.tex
In this section we describe the LMTO basis-set used to  calculate
 the excited states of solids. We discuss the 
basis used for an all electron calculation where the potential is not
supposed to be spherically symmetric  nor of muffin-tin type. 
The use of  a general potential makes the study of open structures possible
without having to resort to  the so-called ``empty-sphere''
approximation. To define the basis-set, we divide the 
space into non overlapping spheres called ``muffin-tin'' 
spheres and a region between these spheres which we call interstitial region.
Inside the muffin-tin spheres the Schroedinger equation is  solved at 
a fixed energy for each angular momentum $\ell$ and variational parameter
$\kappa$ (which is  defined later). The linearization amounts to the 
use of a linear combination of the solution $\phi_{\ell}(e,r)$ of the 
Schroedinger equation 
for a fixed energy and its energy derivative  $\dot\phi_{\ell}(e,r)$ 
inside the muffin-tin spheres. These linear combinations matche
continuously and differentiably  to
an envelop function (spherical function) in the interstitial region.
The Bloch wave function in the interstitial region is given by 
a linear combination of these Hankel functions centered at each site:

\begin{eqnarray}
%\eqalignno{
  \psi_i(\bvc k,\bvc r) =
      \sum_R e^{i\bvc k\cdot\bvc R}
      \K_{L_{i}}\bigl(\kappa_{i},\bvc r\sminus\boldtau_{i}\sminus\bvc R\bigr)
\label{planewave}
\end{eqnarray}
\noindent 
where $i$ stands for the number of the basis function quantum numbers (these 
numbers are  $\{\tau,L,\kappa, \{e_{\ell t}\}\}$), where $\tau$ is
the site number, $L =(\ell,m)$ groups the two angular quantum numbers, and
$e_{\ell t}$ is the linearization energy for a particular atom type $t$
and angular momentum number $\ell$. The envelop functions  are defined as
$\K_{L}(\kappa,\bvc r) \equiv \K_{\ell}(\kappa,r) \Y_{L}(\hat\bvc r)$. 

\begin{eqnarray}
\Y_{\ell m}(\hat\bvc r) & \equiv & i^{\ell} Y_{\ell m}(\hat\bvc r)  \\
 \K_{\ell}(\kappa,r)&  \equiv&  - \kappa^{\ell+1}
        \left\{ \begin{array}{ll}
             n_{\ell}(\kappa r),\  & \textrm{if $\kappa^2 > 0$} \\
             n_{\ell}(\kappa r) - i j_{\ell}(\kappa r),\  & 
                         \textrm{if $\kappa^2 < 0$,\ 
                $(\kappa = i|\kappa|)$} 
                \end{array} \right.  \\
 \J_{\ell}(\kappa,r) & \equiv & \kappa^{-\ell} j_{\ell}(\kappa r)
\end{eqnarray}

Here $n_{\ell}$ is the Neumann function and $j_{\ell}$ Bessel function for
the angular momentum $\ell$, and $Y_{\ell m}$ are the spherical
harmonics. 

To get the differentiability of the wave-function at the boundary of the
muffin-tin spheres, we write the envelope function inside the muffin-tin
spheres. The envelope function for a muffin-tin sphere $\tau^\prime$ is 
given by:

\begin{eqnarray}
      \sum_R && e^{i\bvc k\cdot\bvc R} \K_{L}\bigl(  
                  \kappa,\bvc r\sminus\boldtau\sminus\bvc R\bigr)
\Big|_{r_{\tau
\Pr}<S_{\tau\Pr}} \\
      & = &
      \sum_{L\Pr} \Y_{L\Pr}(\hat\bvc r_{\tau\Pr})
      \Bigl(
            \K_{\ell\Pr}(\kappa,r_{\tau}) \delta(\tau,\tau\Pr)
\delta(L,L\Pr)  \\ \nonumber
& +  & \J_{\ell\Pr}(\kappa,r_{\tau}) B_{L\Pr,L}(\boldtau\Pr\sminus\boldtau,
\kappa,\bvc k)
      \Bigr)
\end{eqnarray}

To produce  smooth basis functions we require that the basis function is 
differentiable  at the boundary of each muffin-tin sphere, i.e.,  
that  a linear  combination of $\phi$ and $\dot\phi$ matches continuously
and differentiably
$\K$ and $\J$ at the boundary of the parent sphere and other spheres,
respectively. Using these  matching conditions at the muffin-tin spheres, 
the Bloch
wave function inside a muffin-tin sphere $\boldtau$ of the unit cell at the
origin is given by \cite{jwarticle}:
 
\begin{equation}
      \psi_i(\bvc k,\bvc r) \Big|_{r_{\tau}<S_{\tau}}
      =
      \sum_{L}
      \Y_{L}(\CD_{\tau}\hat\bvc r_{\tau})
      U_{\ell}(e_{\ell t i},r_{\tau})
      \Omega(\ell t,e_{\ell t i} \kappa_i) 
      \B_{L,L_{i}}(\boldtau\sminus\boldtau_{i},\kappa_i,\bvc k)  
\label{wavemuffin}  
\end{equation}

\noindent 
where

\noindent 
\begin{equation}
      U_{\ell}(e,r) \equiv \left(\matrix{\phi_{\ell}(e,r),
\dot\phi_{\ell}(e,r)}
\right)
\end{equation}

\begin{equation}
      \Omega(\ell t,e\kappa)
      \equiv
      S_{\tau}^2
      \left(\begin{array}{cc}
       - W(\K,\dot\phi) & - W(\J,\dot\phi) \\
         W(\K,\phi)     &   W(\J,\phi)     
           \end{array}\right)
      \ \ \ (W(f,g) \equiv fg\Pr - f\Pr g)
\end{equation}

\begin{equation}
      \B_{L,L_{i}}(\boldtau\sminus\boldtau_{i},\kappa_i,\bvc k)
      \equiv
      \left(\begin{array}{ll}  
        \delta(\tau,\tau_{i}) \delta(L,L_{i}) \\
         B_{L,L_{i}}(\boldtau\sminus\boldtau_{i},\kappa_i,\bvc k) 
            \end{array}\right)
\end{equation}

To add the spin dependence to the basis-set, the Bloch wave function
is multiplied by the eigenvector of the Pauli spin operator $\eta_{\pm
1}$:

$$ \psi_{\sigma}(\bvc k,\bvc r) = \psi(\bvc k,\bvc r) \eta_{\sigma} $$

Such that: 

$$ \hat\bvc n\cdot \boldsigma \eta_{\pm 1} = (\pm 1) \eta_{\pm 1} $$

\noindent
where $\eta$ is the quantization axis chosen in advance.
\section{Dielectric function}
%input dielectric.tex
\noindent
\subsection{\bf Dynamical Dielectric Function}

Here we give a review of the determination
of the dielectric response  of a semiconductor due to the
application of an electric field. We expend  the description of 
our published work \cite{aw} by giving more details concerning the calculation 
of the momentum matrix elements. 

An electromagnetic field of frequency $\omega$, and  a wave
vector $\bf q+G$ interacting  with atoms in a crystal produces a 
response of frequency $\omega$ and
a wave vector  $\bf q+G^\prime$  (${ \bf G} $ and ${\bf G^\prime}$ being
reciprocal lattice vectors). The microscopic field of wave vector
 $ {\bf q+G^\prime}$ is produced by the umklapp processes as a result of
the applied field  $E_0({\bf q + G}, \omega)$

%=====================================================================
\begin{equation}
 E_0({\bf q + G}, \omega) =
   \sum_{\bf G^\prime} \epsilon_{\bf G, G^\prime} ({\bf q}, \omega) E({\bf
q + G^\prime}, \omega)
\label{field}
\end{equation}
%=====================================================================
\noindent
where  $ E({\bf q + G}, \omega)$ is the total field  producing the
non-diagonal elements in the microscopic dielectric function
$\epsilon_{\bf G, G^\prime} ({\bf q}, \omega)$.
The microscopic dielectric function in the random phase approximation 
is given by \cite{adlerwiser}:

%=====================================================================
\begin{eqnarray}
\epsilon_{\bf G, G^\prime} ({\bf q}, \omega)  &=& 
\delta_{\bf G, G^\prime} - { 8 \pi e^2 \over
{\Omega {|{\bf q + G}||{\bf q + G^\prime}|}}}\\\nonumber 
&& \times \sum_{{\bf k}, n,n^\prime} {{f_{n^\prime,{\bf k+q}} -
f_{n,{\bf k}}} \over {E_{n^\prime,{\bf k+q}} - E_{n,{\bf k}}-\hbar \omega
+ i\delta}} 
\\ \nonumber 
&& \langle n^\prime, {\bf k+q}| e^{i ({\bf q + G}) {\bf r}} |
n,{\bf k} \rangle
\langle n,{\bf k}| e^{-i ({\bf q + G^\prime}) {\bf r}}
 | n^\prime,{\bf k + q} \rangle
\label{adler}
\end{eqnarray}
%=====================================================================

Here $n$ and  $n^\prime$ are the band indexes, $f_{n,{\bf k}}$ is the
zero temperature 
Fermi distribution, and $\Omega$ is the cell volume.
The energies  $E_{n,{\bf k}}$ and the the crystal
wave function  $ | n,{\bf k} \rangle$ are produced for each band index $n$
and  for each wave vector  ${\bf k}$ in the Brillouin zone.

The macroscopic dielectric function in the   infinite wave
length limit is given by the inversion of the microscopic dielectric
function:

%=====================================================================
\begin{eqnarray}
 \epsilon (\omega) & =&  \lim_{{\bf q} \to {\bf 0}} { 1 \over
[\epsilon_{\bf G, G^\prime}^{-1} ({\bf q}, \omega)]_{\bf 0,0} } \\
\nonumber
& =& \epsilon_{0,0}(\omega) -  \lim_{{\bf q} \to {\bf 0}}
\sum_{{\bf G, G^\prime \ne 0}} \epsilon_{0,\bf G}({\bf q}, \omega)
T^{-1}_{\bf G, G^\prime} ({\bf q}, \omega) \epsilon_{\bf G^\prime,0}({\bf
q},
\omega) \;
\label{localfield}
\end{eqnarray}
%=====================================================================

Where $T^{-1}_{\bf G, G^\prime}$ is the inverse matrix of $T_{\bf G,
G^\prime}$ containing the elements $\epsilon_{\bf G, G^\prime}$ with
${\bf G}$ and $\bf  G^\prime \ne \bf 0$. The first term of this
equation is the interband contribution to the macroscopic dielectric
function  and the second term represent the local-field correction to
$\epsilon$. The most recent  ab-initio pseudopotential calculation found
that the local-field effect reduces the static dielectric function by at
most
5\% \cite{zachary}. Previous calculations with the same method have also
found a decrease of $\epsilon_{\infty}$ by about the same percentage
\cite{hl,baroni}. 
For insulators the dipole approximation of  the imaginary part of the
first term of equation (\ref{localfield}) is given by \cite{eh59}:

%=====================================================================
\begin{equation}
\epsilon_2(\omega)={   e^2 \over 3\omega^2\pi}\sum_{n,n^\prime}
\int  d{\bf k} | \langle n, {\bf k}| {\bf v}  | n^\prime, {\bf k}\rangle
|^2
f_{n,{\bf k}} (1-f_{{n^\prime},{\bf k}} ) \delta (e_{{\bf k},n^\prime,n}
 -\hbar\omega)\;,
\label{ehrenreich}
\end{equation}
%=====================================================================

\noindent
Here  ${\bf v}$ is the velocity operator, and in the LDA ${\bf v} = {\bf
p }/m$ ($\bf p $ being the momentum operator), and
where  $e_{{\bf k},n,n^\prime} = E_{n^\prime,{\bf k}} -
E_{{n},{\bf k}}$.
The matrix elements  $\langle n {\bf k}| {\bf p} |n^\prime {\bf
k}\rangle$
are calculated for each projection $p_j={\hbar \over i}\partial_j$,
$j= x$ or $y$ and $z$, with the wave function $| n {\bf k}>$ expressed in
terms
 of the full-potential LMTO  crystal wave function described by
equations (\ref{wavemuffin}) and (\ref{planewave}).  The {\bf k}-space
integration is performed using
the tetrahedron method \cite{ja72} with a large number of 
 irreducible {\bf k} points the Brillouin zone.
The irreducible {\bf k}-points are obtained from a shifted  {\bf k}-space
grid from the high symmetry planes and $\Gamma$ point by a half step in 
each of the $k_x$,
$k_y$, and $k_z$ directions. This scheme  produces highly accurate
integration
in the Brillouin zone by avoiding high symmetry points.

\subsection{\bf Momentum Matrix Elements}
To calculate these matrix elements we first defined a tensor operator of
order
one out of the momentum operator
 $\nabla_0 = \nabla_z = { \partial \over \partial z} \quad\hbox{and}\quad
 \nabla_{\pm 1} = \mp {1 \over \sqrt 2} ({ \partial \over \partial x}
\pm i {\partial \over \partial y} ) $.
The muffin-tin part of the momentum matrix elements is calculated using
the commutator $[ \nabla^2, x_\mu]=2 \nabla_{\mu} $ so that:
%=====================================================================
\begin{eqnarray}
\int_{S_{\tau}} d{\bf r} \phi_{\tau\ell^\prime} (r) & Y_{\ell^\prime
m^\prime}
(\widehat{{\bf r -\tau}})
\nabla_{\mu} \phi_{\tau\ell}(r) Y_{\ell m}(\widehat{{\bf r-\tau}}) =
-{i \over 2} G^{1\mu}_{\ell m, \ell^\prime, m^\prime}\nonumber  \\
& \int_0^{S_{\tau}} r^2 dr \phi_{\tau\ell^\prime} ({2\over r} {d \over {d
r}} r
+ { {{\ell (\ell +1)} - {\ell^\prime (\ell^\prime +1)} } \over r })
\phi_{\tau\ell} (r)
\label{pmuffin}
\end{eqnarray}
%=====================================================================
where $G^{1\mu}_{\ell m, \ell^\prime, m^\prime}$ are the usual Gaunt
coefficients, and $S_{\tau}$ is the radius of the muffin-tin sphere of
atom $\tau$.
 In the interstitial region the plane-wave
representation of the wave function (see equation ~\ref{planewave})
makes the calculation straightforward, but a
special care has to be taken for the removal of the extra
contribution in the muffin-tin spheres. However,
 we find it much easier and faster to transform the
interstitial matrix elements as an integral over the surface of the
muffin-tin spheres
using the commutation relation of the momentum operator and the
Hamiltonian in the interstitial region. The calculation of the
interstitial momentum matrix elements  is then similar to the
calculation of the interstitial  overlap matrix elements. The
$\kappa = 0$ case has been already derived by Chen using the Korringa,
Kohn and
Rostoker Greens-function method \cite{abchen}. We have tested that both the
plane-wave summation and the surface integration provide the same results.

$$ - \nabla^2 \, {\bf p}\psi  = \kappa^2 {\bf p}\psi $$
A Hankel function can be integrated over a volume
by knowing its integral over the bounding surface:

\begin{eqnarray}
&&  \int_{\I}  d^3r  \nabla\left(\Conjg{\psi}_1 \nabla p_i \psi_2 
                        - 
       \left(\nabla \Conjg{\psi}_1\right) p_i \psi_2 \right)\nonumber \\
& = & 
 \left(\kappa_1^2 - \kappa_2^2\right) \int_{\I} d^3r \Conjg{\psi}_1 p_i \psi_2
\end{eqnarray}

The surface of the interstitial consists of the exterior of the muffin-tin 
spheres and the unit cell boundary.

Over the surface of the muffin tins:  the surface area is $S^2 d\Omega$ 
and the normal to the sphere points inward

\begin{eqnarray}
    &&  \left(  \kappa_1^2  - \kappa_2^2 \right) \int_{\I} d^3r\ 
\Conjg{\psi}_1 p_i  \psi_2  =    \\ \nonumber 
     &&  - \sum_{\tau} S_\tau^2
      \int d^S\ 
                   \left( 
                   \Conjg{\psi}_1 {\partial\over\partial r} p_i\psi_2 
                        - 
          \left( {\partial\over\partial r}\Conjg{\psi}_1 \right) p_i\psi_2 
\right)
\end{eqnarray}
At a muffin-tin sphere boundary $S_\tau$ the Bloch wave function is 
given by:

\begin{eqnarray}
      &\psi_i & (\bvc k,\bvc r) 
      \Big|_{S_\tau}
      = 
      \sum_R e^{{\rm{i}}\bvc k\cdot\bvc R} 
      \K_{L_{i}}\bigl(\kappa_{i},\bvc r\sminus\tau_{i}\sminus\bvc R\bigr)
      \Big|_{S_\tau} \\ \nonumber 
      & =  & 
      \sum_{\ell m} \Y_{\ell m}(\hat\bvc r)
      \RK_{\ell}(\kappa_i,S)
      \RB_{\ell m,\ell_i m_i}(\boldtau\sminus\boldtau_i,\kappa_i,\bvc k)
\end{eqnarray}

where $\RB_{\ell m,\ell_i m_i}(\boldtau\sminus\boldtau_i,\kappa_i,\bvc k)
= \left(\matrix{
            \delta(\tau,\tau_i)
            \delta(\ell,\ell_i)
            \delta(m,m_i)
            \cr
            B_{\ell m,\ell_i m_i}(\boldtau\sminus\boldtau_i,\kappa_i,\bvc
k)
            \cr
      }\right)
$ and 
$\RK = \left(\matrix{ \K \cr \J \cr }\right) $

Let W denotes the Wronskian   $W(f,g) = f g^\prime -f^\prime g$

We define then

\begin{equation}
S^2 W_0 = S^2 W(\RK_\ell^T(\kappa ), \RK_\ell(\kappa )) = \left(
\begin{array}{cc}
0 & 1 \\
-1 & 0       
\end {array} \right)  \nonumber
\end{equation}

and 

\begin{equation}
w_{\tau \ell} {\kappa_1, \kappa_2} = S_\tau \frac{ W(\RK_\ell^T(\kappa )),
\RK_\ell(\kappa) - W_0}{\kappa_1^2 - \kappa_2^2} \nonumber
\end{equation}

\begin{eqnarray}
 {\bf p}\psi_i |_\tau & = &
       \sum_{\mu} \hat\bvc e_{\mu}\  \sum_{\ell m} \biggl[ \RK_{\ell-1 m-\mu}
      \left(\begin{array}{cc}
            \kappa^2_i  & 0 \\
             0          & 1  
      \end{array} \right)
      \gauntcof {\ell\sminus1} {m\sminus\mu} \ell m 1 \mu 
\nonumber  \\
&& \sminus  \RK_{\ell+1 m - \mu}
\left(\begin{array}{cc}
            1  & 0 \\
            0 & \kappa^2_i  
      \end{array} \right)
      \gauntcof {\ell\splus1} {m\sminus\mu} \ell m 1 \mu
      \biggr] \nonumber  \\
&& \RB_{\ell m,\ell_i m_i}(\boldtau\sminus\boldtau_i,\kappa_i,\bvc k)
      \Bigr)
\end{eqnarray}

then

\begin{eqnarray}
&\langle & \psi_f  {\bf p}\psi_i\rangle_\tau  =  \\ \nonumber 
&&    \sum_{\tau}\sum_{\mu} \hat\bvc e_{\mu}\  \sum_{\ell m} 
    \biggl[ \RB_{\ell-1 m-\mu,\ell_f m_f}
(\boldtau\sminus\boldtau_f,\kappa_f,\bvc k) w_{\tau \ell-1}(\kappa_f,\kappa_i) 
\\ \nonumber 
&&\times  \left(\begin{array}{cc}
  \kappa^2_i  & 0 \\
            0 & 1  
      \end{array} \right)
      \gauntcof {\ell\sminus1} {m\sminus\mu} \ell m 1 \mu  \\ \nonumber
&& \sminus  \RB_{\ell+1 m-\mu,\ell_f m_f}
(\boldtau\sminus\boldtau_f,\kappa_f,\bvc k) w_{\tau \ell+1}(\kappa_f,\kappa_i) 
\\ \nonumber 
&&\times \left(\begin{array}{cc}
            1  & 0 \\
            0 & \kappa^2_i  
      \end{array} \right)
      \gauntcof {\ell\splus1} {m\sminus\mu} \ell m 1 \mu
      \biggr]  \\ \nonumber 
&& \RB_{\ell m,\ell_i m_i}(\boldtau\sminus\boldtau_i,\kappa_i,\bvc k)
      \Bigr)  + \Delta(f, i,\kappa_i)
\end{eqnarray}

where 

\begin{eqnarray}
&&(\kappa_f^2 - \kappa_i^2) \Delta(f, i,\kappa_i) =  
    \sum_{\mu} \hat\bvc e_{\mu}(\tau_i) \biggl[ \\ \nonumber  
&&+ B^{\star}_{\ell_i+1 m_i-\mu,\ell_f m_f}
(\boldtau_i\sminus\boldtau_f,\kappa_f,\bvc k) 
   \gauntcof {\ell_i\splus1} {m_i\sminus\mu} {\ell_i} {m_i} 1 \mu
\kappa^2 \\ \nonumber 
&&   - B^{\star}_{\ell_i-1 m_i-\mu,\ell_f m_f}
       (\boldtau_i\sminus\boldtau_f,\kappa_f,\bvc k) 
      \gauntcof {\ell_i \sminus1} {m_i\sminus\mu} {\ell_i} {m_i} 1 \mu \kappa^2
       \biggr] \nonumber \\ 
&&  + \sum_{\mu} \hat\bvc e_{\mu}(\tau_f) \biggl[  \\ \nonumber 
&& +  B_{\ell_f+1 m_i+\mu,\ell_f m_f}
(\boldtau_f\sminus\boldtau_i,\kappa_i,\bvc k) 
   \gauntcof {\ell_f} {m_i} {\ell_f\splus1} {m_f\splus\mu} 1 \mu \\ \nonumber 
&&   - B_{\ell_f-1 m_f-\mu,\ell_i m_i}
(\boldtau_f\sminus\boldtau_i,\kappa_i,\bvc k) 
      \gauntcof {\ell_f} {m_f} {\ell_f\sminus1} {m_f\splus\mu} 1 \mu \kappa^2
       \biggr] 
\end{eqnarray}

\subsection{\bf Velocity Operator and Sum Rules}

Equation (\ref{ehrenreich}) can not be used directly to determine the 
optical properties
of semiconductors, when the GW approximation or the scissors operator is 
used to determine the electronic structure. The velocity 
operator should be obtained from the effective momentum operator ${\bf
p^{eff}}$ which is calculated using the self-energy operator, 
$\Sigma({\bf r, \bf p})$, of the system \cite{delsole}:
%=====================================================================
\begin{equation}
{\bf v} = {\bf p^{eff}} /  m  = {\bf p} / m + \partial
\Sigma({\bf r, \bf p}) / \partial {\bf p}
\label{velocity}
\end{equation}
%=====================================================================
 GW calculations show  that  the quasiparticle wave function is 
almost equals  to the LDA wave function \cite{hl,gss}.
Based on this assumption,
%  Del Sole and Girlanda show that the effective momentum operator 
% ${\bf p^{eff}}$ can be written in terms of the momentum operator ${\bf
% p}$ as follows \cite{delsole}:
% %=====================================================================
% \begin{equation}
% \langle n^\prime, {\bf k}|{\bf p^{eff}} | n, {\bf k}\rangle  =
%  \langle n^\prime, {\bf k}| {\bf p}  | n, {\bf k}\rangle  
% {e^{QP}_{{\bf k},n^\prime,n} / e_{{\bf k},n^\prime,n}},
% \label{pqp}
% \end{equation}
%=====================================================================
% where $e^{QP}_{{\bf k},n^\prime,n} = 
% E^{QP}_{n^\prime,{\bf k}} -  E^{QP}_{{n},{\bf k}}$
% is the difference between the quasiparticle energy $E^{QP}_{n^\prime,{\bf
% k}}$ of the unoccupied
% state $|n^\prime, \bf k \rangle$ and the occupied state $|n, \bf k\rangle$. 
% By substituting
% Equation ~\ref{pqp} into equation ~\ref{ehrenreich}, 
it can be easily shown \cite{delsole} that in the case 
of the scissors operator, where all the empty states are shifted 
rigidly by  a constant energy $\Delta$, the imaginary part of the dielectric 
function is a simple energy shift of the LDA dielectric 
function towards the high energies by an amount $\Delta$, i.e., 
$\epsilon_2^{QP}(\omega) = \epsilon^{\rm LDA}_2(\omega -\Delta/\hbar)$. The 
real part of the dielectric function is then obtained from the 
shifted $\epsilon_2$  using Kramers-Kronig relations. 
The expression of $\epsilon_{\infty}^{QP}$ is given by:  
%=====================================================================
\begin{equation}
\epsilon^{QP}_{\infty}= 1 + {2e^2 \over   3\omega^2\pi^2}\sum_{n,n^\prime}
\int  d{\bf k}f_{n,{\bf k}} (1-f_{{n^\prime},{\bf k}} )
{ | \langle n, {\bf k}| {\bf p}  | n^\prime, {\bf k}\rangle |^2 \over 
 (e_{{\bf k},n^\prime,n} +\Delta) e_{{\bf k},n^\prime,n}^2}\;,
\label{estaticqp}
\end{equation}
%=====================================================================
$\epsilon^{QP}_{\infty}$ is very similar to $\epsilon^{\rm LDA}_{\infty}$
except that one of the interband gap $e_{{\bf k},n^\prime,n}$ is
substituted by the QP interband gap $e_{{\bf k},n^\prime,n} +\Delta$. 

To test for the
accuracy of the calculation within the LDA the f-sum rule:
%=====================================================================
\begin{equation}
{2 \over 3 m n_v} \sum_{\bf k} \sum_{n,n^\prime} f_{n,{\bf k}}
(1-f_{{n^\prime},{\bf k}} ){|\langle n, 
{\bf k}|{\bf p} | n^\prime, {\bf k}\rangle|^2 \over e_{{\bf k},n^\prime,n}} 
= 1, 
\label{fsumrule}
\end{equation}
%=====================================================================
where $n_v$ is the number of valence bands, should be always  checked 
to ensure the accuracy of the  calculations.

It is easily seen that the dielectric function $\epsilon_2^{QP}$ 
calculated using
the scissors-operator shift does not satisfy the 
sum rule ($\omega_P$ is the free-electron plasmon frequency):
%=====================================================================
\begin{equation}
\int_0^{\infty} \omega
\epsilon_2 (\omega ) d\omega  = {\pi \over 2} \omega_P^2
\label{intsumrule}
\end{equation}
%=====================================================================
because (i)
$\epsilon_2^{\rm LDA}$ satisfies this rule, and (ii) $\epsilon_2^{QP}$ is
obtained by a simple shift of $\epsilon_2^{LDA}$ by the scissors-operator
$\Delta$ towards higher energies. 
%Using the expression of the quasiparticle dielectric function 
%in the scissors-operator shift approximation we show 
%that $\epsilon_2^{QP}$ satisfy the following integral sum rule:
%%=====================================================================
%\begin{equation}
%\int_0^{\infty} \omega
%\epsilon_2^{QP} (\omega ) d\omega  = {\pi\over 2} {\omega^{\prime}_P}^2 
%\label{intsumruleqp}
%\end{equation}
%=====================================================================
%where   ${\omega^{\prime}_P}^2 = \omega_P^2 + 
%{2 e^2 \Delta \over {3\pi^2 m^2}} \sum_{n, n^\prime} 
%\int d{\bf k}{ | \langle n, {\bf k}| {\bf p}  | n^\prime, {\bf k}\rangle |^2 / 
%e_{{\bf k},n^\prime,n}^2} f_{n,{\bf k}} (1-f_{{n^\prime},{\bf k}} )$. 
%We recover the usual sum rule when  $\Delta $ is equal to zero. 
The non
simultaneous satisfaction of both the f-sum rule and the integral sum
rule within the scissors approximation shows the  
limitation of this approximation. While the scissors operator
approximation describes nicely the low lying excited states, which is seen
in the good determination of the static dielectric function and the low
energy structures, i.e. $E_1$ and $E_2$, in the imaginary part of the dielectric
function, it seems to fail for the description of the higher excited states. 
This is not surprising because  the higher excited states which are free
electrons like are most probably well described within LDA and  need no 
scissors-operator shift.  
This is supported by the fact that the the energy-loss 
function, -Im$\epsilon^{-1}$,  within the
LDA has it maximum roughly at  the free electron 
plasmon frequency whereas within the scissors approximation its maximum is
shifted to higher energies. 
For our purpose the scissors-operator
shift remains a good approximation for the description of the low-lying 
excited states of semiconductors and their optical properties.

\section{Applications}
\subsection{Optical Properties}
%input optics.tex
We have used our FP-LMTO method and the formalism outlined above to
calculate the optical properties of materials \cite{logoetal,aw,delin,ahuja}. 
In general our  results are often in good agreement with the experimental 
results. For semiconductors, however,   good agreement with
experiment is only achieved when the so called scissors-operator shift 
is used. Figure 1 presents
our relativistic calculation of the imaginary part of the dielectric 
function of GaAs compared  to the experimental results of Ref. \cite{aspnes}.       
\begin{figure}[t]
\includegraphics[width=.7\textwidth]{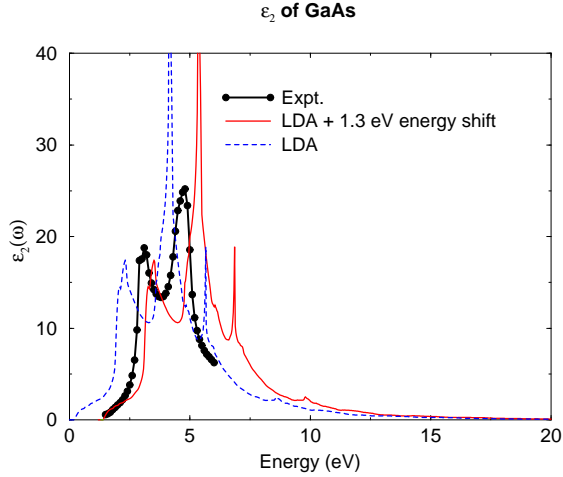}
\caption{Calculated Imaginary part  of the dielectric function 
of  GaAs at the experimental equilibrium volume both within LDA and shifted
by 1.3 eV, compared with the experimental results of
Ref. \protect\cite{aspnes}. The
experimental $E_1$ is only slightly underestimated while  $E_2$ is
overestimated. Notice that the shifted dielectric function by  1.3 eV, 
which produces the correct band gap, overestimates the peak positions by
about 0.3 eV. Excitonic effect should shift these peaks to lower values in
agreement with experiment.
}
\label{e2gaas}
\end{figure}
The LDA relativistic results underestimates the band gap by about 1.3 eV.
When the imaginary part of the dielectric function is shifted to higher
energies by 1.3 eV the results the $E_1$ and $E_2$ peaks are overestimated
in our calculation. One needs to shift the spectrum by less than the band
gap as done in Ref. \cite{aw} to produce good agreement with experiment.
It seems then that the optical band gap is less than the band energy gap (1.5
eV). The optical band gap is produced by interband transitions to the 
low lying conduction states. Excitonic effects are therefore important and
are responsible for the reduction of the energy gap of semiconductors. 
It is interesting to notice though that the static dielectric function
are in good agreement with experiment for GaAs, Si, and Ge when the
shift correspond to the energy band gap obtained from photoluminescence
\cite{zachary,aw}.

\begin{figure}[t]
\includegraphics[width=.8\textwidth]{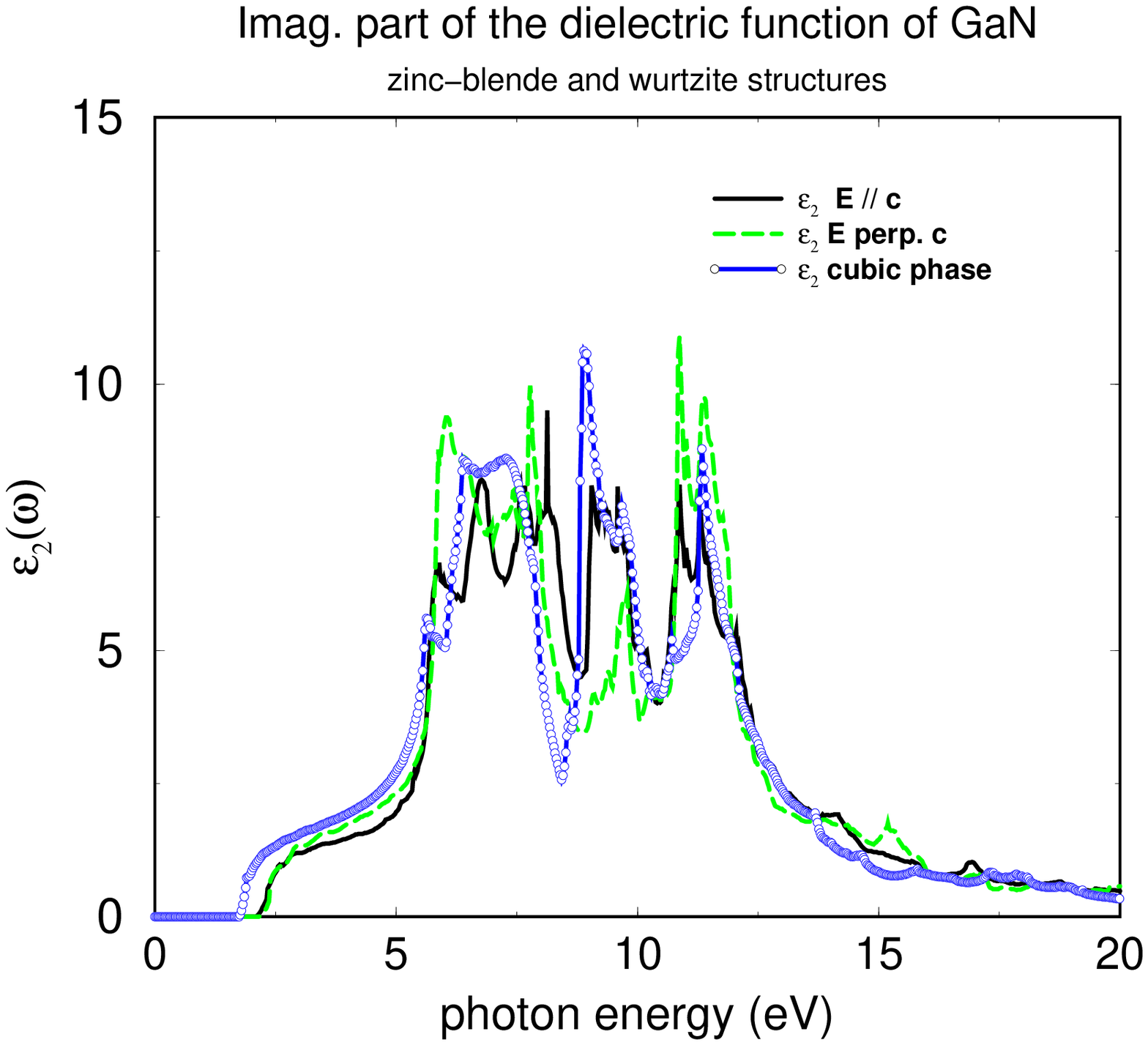}
\includegraphics[width=.8\textwidth]{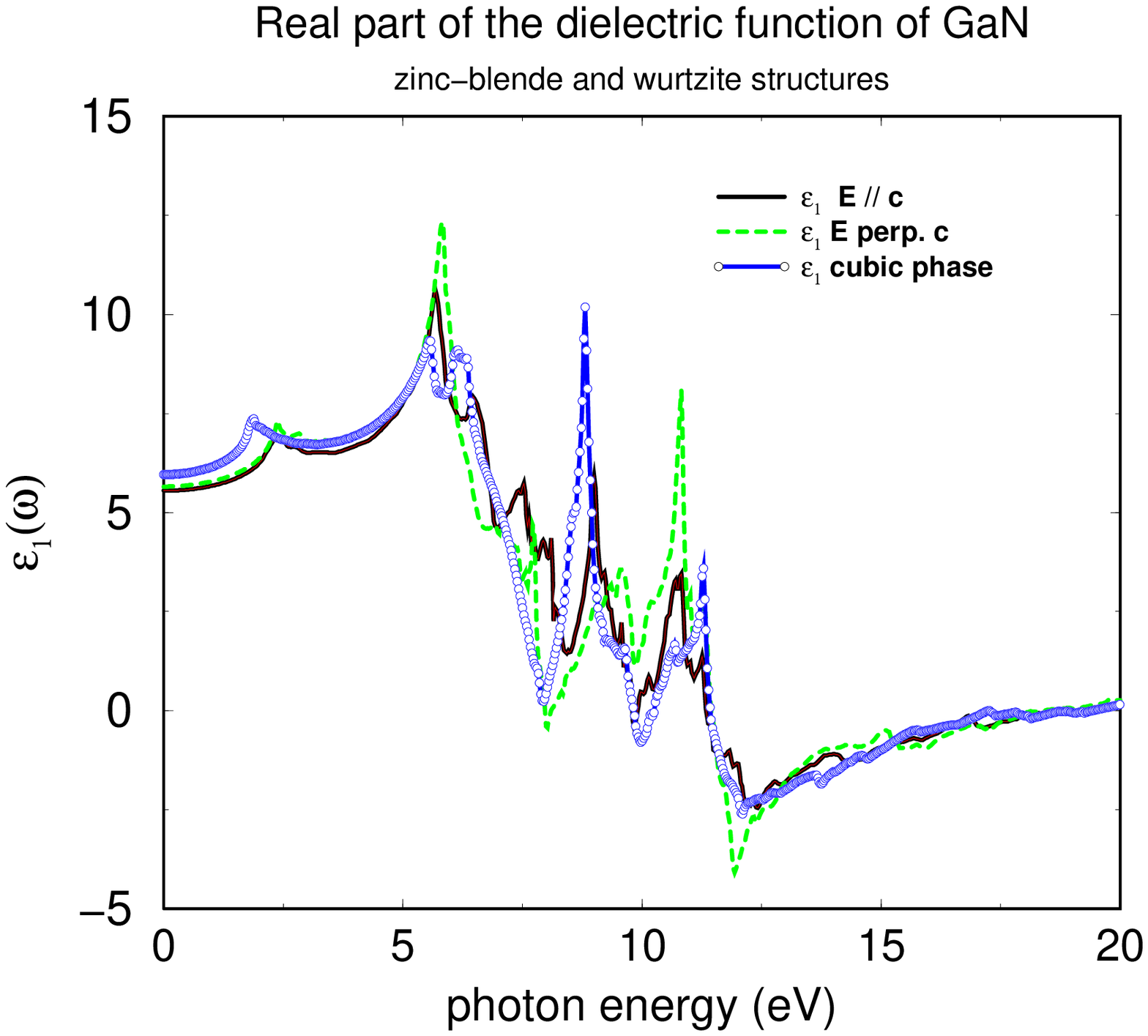}
\caption{Calculated imaginary and real parts  of the dielectric function of GaN in 
its cubic and wurtzite forms. The LDA band gap of the cubic phase is 1.8 eV and
the wurtzite phase is 2.2 eV. 
}

\label{gan}
\end{figure}
More interesting are the wide band-gap materials where the LDA
calculated static dielectric function is in good agreement with
experiment despite that the band gap is still underestimated by
LDA. Correcting the band gap using the scissors operator makes 
the static dielectric much small than
the measured value. As an example of wide gap material, we present
in Figure 2 and 3 the imaginary part of the dielectric function of GaN
for the cubic (3)  and wurtzite structure (B4).

\begin{table}
\caption{
Calculated static dielectric function $\epsilon_\infty$ for GaN compared to 
pseudopotential (PP) results and experiment. For the wurtzite structure
we have calculated   $\epsilon_\infty^\Vert$ for a polarization parallel 
to the $xy$ plan  and $\epsilon_\infty^\bot$ which is perpendicular.
}
\renewcommand{\arraystretch}{1.1}
\setlength\tabcolsep{5pt}
\begin{tabular}{llll}
\hline\noalign{\smallskip}
 & zinc-blende  & wurtzite &  \\ 
 & $\epsilon_\infty$  & $\epsilon_\infty^\Vert$ & $\epsilon_\infty^\bot$  \\ 
\hline 
PP &  5.74 & 5.48 &5.60    \\
Present work & 5.96 & 5.54 & 5.65 \\
Expt.  &  & 5.35 & 5.35$\pm$0.2 \\
\hline
\hline
\end{tabular}
\label{epsilon0}
\end{table}

Table I shows that our LDA dielectric constant calculations are in
agreement with available experimental results and the pseudo-potential
(PP) results \cite{chen} including local-field effects (an error  about our
calculation is reported  in Ref.
\cite{chen}; our value for $\epsilon_\infty^\Vert$ is not 4.48 but 5.54
and the PP value should then be   4.48).      
 It is interesting to notice that    static dielectric
is in good agreement for the for all the nitrides \cite{chen} while 
the band gap is underestimated. The scissors-operator shift fails to 
explain the static dielectric function of large gap semiconductor. 
Recently, both local-field 
effects and electron-hole interaction were included on an ab-initio
computation of the dielectric function of few 
semiconductors \cite{alb,benedict} by extending the semi-empirical
Hanke and coworkers approach \cite{strinati,hanke}
which is based on the solution of the Bethe-Salpeter equation \cite{hanke}.
The excitonic effects seem to improve significantly the agreement between
theory and experiment. However for large band-gap semiconductors, such
as diamond, the inclusion of the excitonic effects seem to underestimate 
the optical band gap by as much as 1 eV \cite{benedict}. 
It is not clear from these calculations
whether the static dielectric function for wide-band gap semiconductors 
is improved when excitonic effects are included. More theoretical work 
along these lines is needed to fully understand the dielectric function
of wide-gap semiconductors. 

\subsection{Magnetic Circular Magnetic Dichroism}
%input mxd.tex
 X-ray absorption spectroscopy (XAS)
probes selectively each core orbital of each
atomic species in a material. Two decades ago  the theoretical work of
Erskine
and Stern show that the x-ray absorption could be used to determine the
x-ray magnetic circular  dichroism (XMCD) in transition metals
when left and right circularly polarized x-ray beams are used
\cite{erskine}. More recently  these ideas were implemented experimentally
and XAS was  used to determine the local
magnetic properties of each magnetic atomic orbital in a  magnetic
compound \cite{laan86,shutz}.  Thus the circular magnetic x-ray dichroism  is
an important tool for the investigation of magnetic materials
\cite{laan86,shutz,carra90,ctchen,carra91,carra92,ebert,guo,brouder,rehr,smith,freeman}, 
 especially through the use of  sum rules for the direct determination of
the local orbital and spin contributions to the total magnetic moment
\cite{carra92}.

Thole and co-workers show that the  circular-magnetic-x-ray  dichroism 
is related to the magnetic moment
of the photo-excited atom when the core electron is excited to the
conduction states that are responsible of the magnetic properties of 
the material.
On the theoretical side, Ebert and his co-workers \cite{ebert,guo}
have developed a fully-relativistic local-spin-density-approximation
approach that was used with success to calculate the XMCD at the K-edge of
Fe, the L$_{3}$-edge of Gadolinium, and Fe and Co multilayers. 
Wu et al used slab
linear augmented plane wave method to study the L$_{2,3}$ XMCD of Fe
\cite{freeman}.  Brouder and co-workers
uses  Multiple-scattering theory to solve the Schr\"{o}dinger
using spherical potentials and spin-orbit coupling as a perturbation
in the final state \cite{brouder}. Recently Ankudinov and Rehr used a
method based on
a non-relativistic treatment of propagation based on high order multiple
scattering theory and spinor-relativistic Dirac-Fock treatment of the
dipole matrix elements to calculate the Fe K edge and Gd L$_3$
edge XMCD \cite{rehr}.

The calculation of the x-ray absorption for  left and right circularly
polarized x-ray beams is implemented within the local-density
approximation (LDA) by means of all-electron full-relativistic
and spin-polarized full-potential linear muffin-tin orbital method (LMTO).
The core electrons are spin-polarized and their electronic states are
obtained by solving the full-Dirac equation, whereas for the valence
electrons the spin-orbit  coupling is  added perturbatively to the
the semi-relativistic Hamiltonian. The total Hamiltonian is then solved
self-consistently.
To calculate the polarization dependent  cross-section
we consider the case where the internal field polarizes the spins along
the magnetization easy axis. With respect to this axis  we defined the
left- and right-circular polarization, which  correspond to the
photon helicity ($+\hbar$) ($-\hbar$) respectively  and the following
dipole
interaction:
$  {\bf \hat  e_{\pm}} {\bf p } = {1\over {\sqrt 2}} (\nabla_x \pm
{\rm i}\nabla_y) $.
The absorption cross-section $\mu_{\pm}$ for left ($+$) and right ($-$)
circular polarized
x-ray calculated at the relativistic j$_\pm$ ($\ell \pm {1\over 2}$) core
level
and  in the dipole approximation is given by:
\begin{equation}
\mu_{\pm} (\omega) = {2\pi \over {\hbar}} \sum_{ m_{j_{\pm}}}
\sum_{n, {\bf k}} \langle j_{\pm}{m_{j_{\pm}}}|  {\bf \hat e_{\pm}}
{\bf p } | {n {\bf k} }\rangle \langle {n {\bf
{\bf k}}}| {\bf p } {\bf \hat e_{\pm}} | j_{\pm} {m_{j_{\pm}}} \rangle
\delta(\omega - E_{n \bf k} + E_{ j_{\pm}} )
\end{equation}
\noindent
using LDA  in conjunction with the
relativistic full-potential LMTO technique.

\begin{figure}[t]
\includegraphics[width=.7\textwidth]{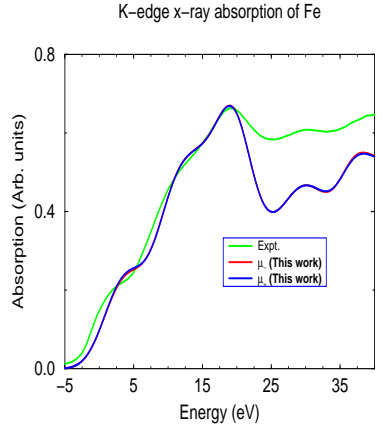}
\caption{Calculated  x-ray absorption at the K-edge of Fe for left and 
right circularly polarized light compared to the experimental spectrum. 
The difference between the two spectra (barely visible on the graph) 
represents the x-ray magnetic circular dichroism.
}
\label{kedgefe}
\end{figure}

Figure 3. represent the K-edge x-ray absorption of Fe, for left and 
right circularly polarized light, compared to the
experimental results. The agreement at low energy with experiment is 
good and start degrading at higher energies above the mean absorption
peak. It is of interest to point out that the magnetic x-ray dichroism 
at the K-edge which is due to the spin polarization and the spin-orbit 
in the final state is very small in the case of Fe. The difference between
the right and left circularly polarization of the light is not even
visible on the graph. However, the x-ray magnetic circular dichroism can
be measured and Figure 4 shows a good agreement of the calculated 
dichroic signal with the experimental results of Sh\"utz \cite{shutz}. 

\begin{figure}[t]
\includegraphics[width=.7\textwidth]{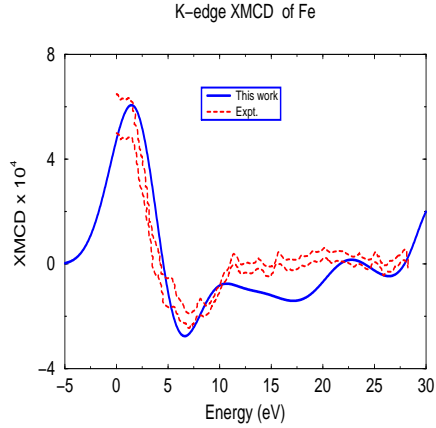}
\caption{Calculated  x-ray magnetic x-ray dichroism at the K-edge f Fe 
compared to the experimental spectrum of Sh\"utz\protect{\cite{shutz}}. 
}
\label{kmxdfe}
\end{figure}

At the L$_{2,3}$ edge of $3d$ transition metals the x-ray magnetic
dichroism is much important because it is meanly due to the presence of
the strong spin-orbit coupling in the initial $2p$ states (in the case of
Fe the spin-orbit splitting between the  $2p_{3/2}$ and $2p_{1/2}$ is 
about 13 eV). In Figure 5 we show the calculated x-ray absorption and XMCD
at the Co in PtCo ordered alloy \cite{grange}.  

\begin{figure}[t]
\includegraphics[width=.7\textwidth]{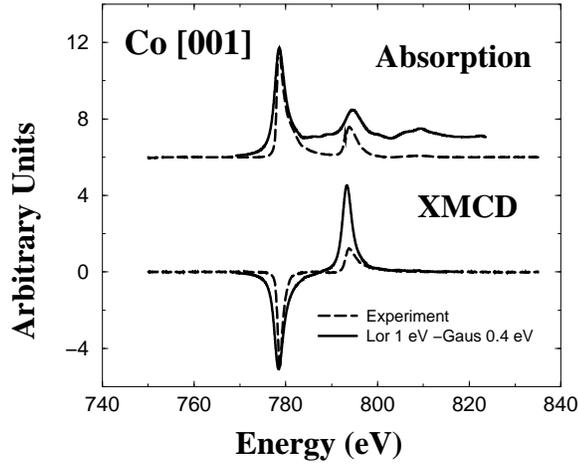}
\caption{Calculated  x-ray absorption and magnetic x-ray dichroism 
at the L$_{2,3}$-edge of Fe compared to the experimental spectrum of 
Grange {\it et al.}\protect{\cite{grange}}. 
}
\label{co}
\end{figure}
To compare the results with experiment we have to take into account the 
effect of the core hole and the experimental resolution. This is done
by convoluting  the calculated spectra by a Lorentzian  
of widths of 0.9 eV  and 1.4 eV for  the   L$_2$ and L$_3$ edges,
respectively,  in addition a  Gaussian broadening of 0.4 eV is added to 
take into account the experimental resolution. 
The calculation of the x-ray magnetic circular dichroic
signal ignoring the electron-hole recombination effect
provides a semi-quantitative  agreement with the experimental spectra.
Hence, we believe that the core hole effect represented here by a 
Lorentzian broadening  plays a significant role in determining the  correct
L$_3 / L_2$ branching ratio for 3$d$ transition metals.
The underestimation of the $L_{2,3}$ branching ratio
remains a challenge for theorists and further theoretical
development along the line proposed by Schwitalla and 
Ebert {\cite{schwit}} is needed to bring the theory at the level of 
the experiment.

For the 4$d$-transition metals, the core hole is deeper, and the agreement
with experiment of the XMCD is satisfactory. Figure 6  shows the
calculated XMCD at the site of Pt of the CoPt ordered alloy. 

\begin{figure}[t]
\includegraphics[width=.7\textwidth]{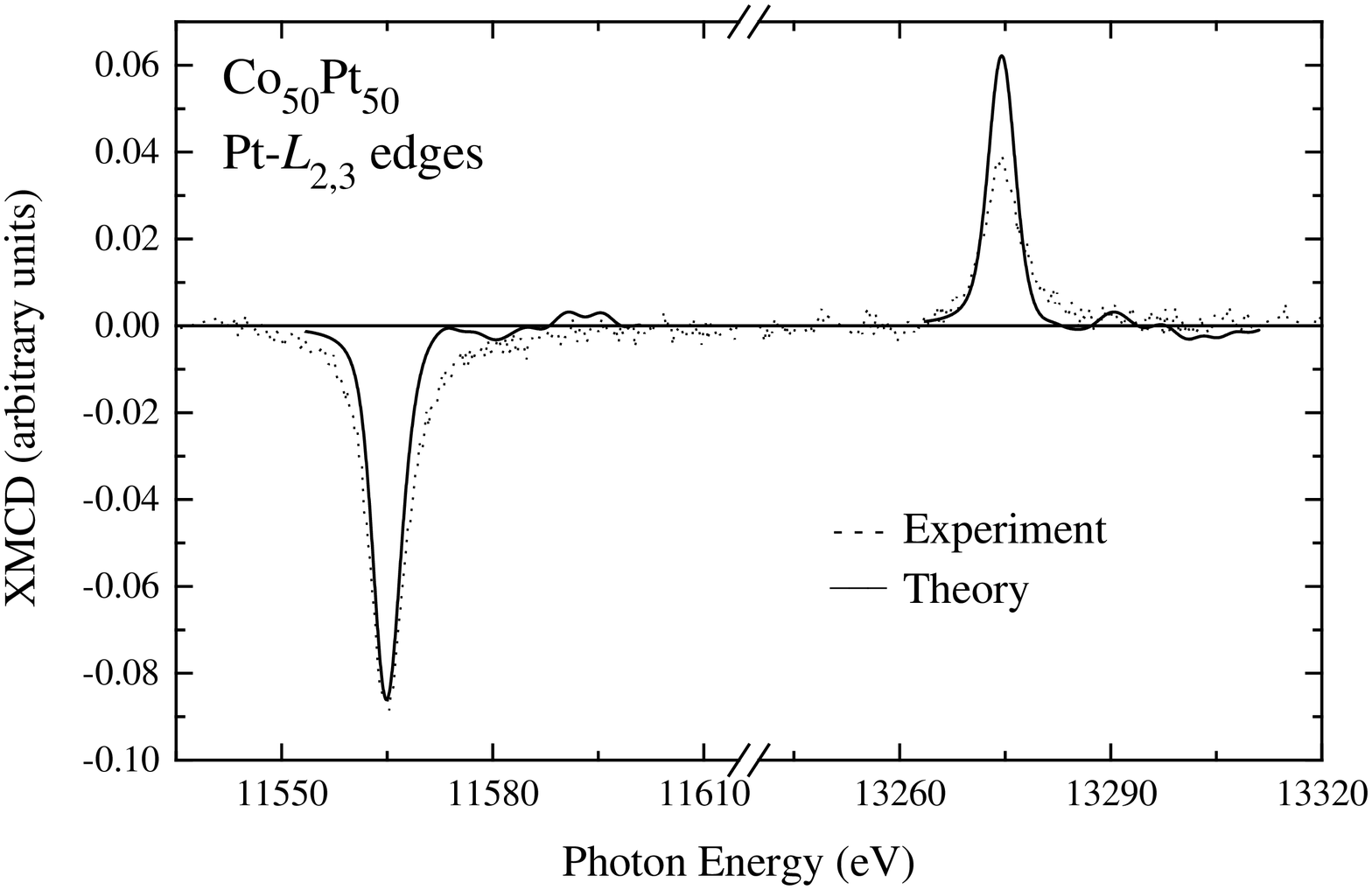}
\caption{Calculated  x-ray  magnetic x-ray dichroism 
at the L$_{2,3}$-edge of Pt compared to the experimental spectrum of 
Grange {\it et al.}\protect{\cite{grange}}. 
}
\label{ltt}
\end{figure}

In contrast  to what is obtained for Co, the results for the Pt site
show a much better agreement with experiment, due to the fact that
the core hole effect is less intense (core hole much deeper than that of
Co).  For the Pt atom we used both a Lorentzian (1 eV) and a Gaussian (1 eV)
to represent the core hole effect and  a Gaussian of 1 eV width for the
experimental resolution.
The experimental and theoretical $L_2$ and $L_3$ edges are separated by a
spin-orbit splitting of  the $2p$ core states of    $1709$ and $1727$ eV 
respectively.
The width of both $L_2$ and $L_3$  edges is comparable to  experiment,
but the calculated $L_2$ edge is much larger.
This produces a calculated  integrated branching ratio
of 1.49 which is much smaller than the
experimental ratio of 2.66. Here again the theory is underestimating
the branching ratio.

\section{Conclusion}
We have reviewed the FP-LMTO method and the implementation of the 
optical properties and x-ray magnetic dichroism within the local density
approximation. We have showed that the momentum matrix elements can be
evaluated as a muffin-tin contribution and a surface term. 
The method has been successfully used to compute the optical properties
of metals \cite{delin,ahuja}, semiconductors \cite{logoetal,aw} and 
magneto-optical properties \cite{delin} of transition metals
alloys, as well as x-ray magnetic circular dichroism
\cite{aww,grange,iosif} with high precision. 

For small-gap semiconductors a scissors-operator shift should be used to 
reproduce
the static and dynamic dielectric function \cite{aw}. 
Excitonic effects seem to be
important in reproducing the correct optical energy gap \cite{alb,benedict}.  
For wide-gap
semiconductors the local-density approximation (LDA)  static dielectric 
function is in good agreement with experiment and no scissors-operator
shift is required despite the underestimation of the band gap by 
LDA \cite{chen}. 

For the computation of the x-ray magnetic circular dichroism the agreement
with experiment is rather 
good \cite{aww,grange,iosif,ebert,guo,brouder,rehr,smith}. 
However, the so called branching ratio is
underestimated by the theory. More theoretical work where the
electron core-hole interaction is taken into account is needed to bring
the theory at the quality level of experiment \cite{schwit}.

Part of this work was done while one of us (M.A) was at Ohio State University and
were supported by NSF, grant number DMR-9520319.  
Supercomputer time was granted by CNUSC (project gem1917) on the
IBM SP2 and by the Universit\'e Louis Pasteur de Strasbourg on the
SGI O2000 supercomputer.

%INDEX%%%%%%%%%%%%%%%%%%%%%%%%%%%%%%%%%%%%%%%%%%%%%%%%%%%%%%%%%%%%%%%
\clearpage
\addcontentsline{toc}{section}{Index}
\flushbottom
\printindex
%%%%%%%%%%%%%%%%%%%%%%%%%%%%%%%%%%%%%%%%%%%%%%%%%%%%%%%%%%%%%%%%%%%%%
\end{document}